\documentclass[12pt]{article}
\usepackage[dvips]{graphicx}

\begin{document}
\vspace{0.2in}
\begin{center}
{Effective Operator Treatment of the Lipkin Model}
\vspace{.3in}

K.J. Abraham$^a$ \& J.P. Vary$^a$
\vspace{.3in}

{\small \it
$^a$ Department of Physics and Astronomy, Iowa State University,
       Ames, Iowa 50011}
\end{center}
\vspace{.5in}
       \begin{center} ABSTRACT  \end{center}
We analyze the Lipkin Model in the strong coupling limit
using effective operator techniques. We present both analytical
and numerical results for low energy effective hamiltonians.
We investigate the reliability of various approximations used to simplify the
nuclear many body problem, such as the cluster approximation. 
We demonstrate, in explicit examples, certain limits to the validity of
the cluster approximation but caution that these limits may be
particular to this model where the interactions are of unlimited range.

\newpage
The effective operator method \cite{leesuz} has been used extensively and
successfully
to obtain the low-lying spectroscopy of complex nuclei
\cite{assorted} with realistic nucleon-nucleon interactions.
Central to the methodology of \cite{leesuz} is the
construction of a similarity transformation which transforms the original
hamiltonian to a new hamiltonian in which a subset of states is decoupled.
It is thus possible to construct an
effective hamiltonian whose eigenvalues are a subset of the eigenvalues
of the full hamiltonian.
When this procedure is implemented the effective hamiltonian generally
will contain all operators
consistent with the symmetries of the bare hamiltonian, even those not
present in the original hamiltonian. In particular, for an
$N$ fermion problem containing just one and two body operators, the effective
hamiltonian can be expected to contain all operators involving up to
$N$ fermions consistent with the symmetries of the bare hamiltonian.
In the nuclear many body problem, it is customary
to make a simplifying approximation such as the cluster approximation
\cite{assorted} where
only one and two body operators in the effective hamiltonian are retained.
Such approximations are also a feature of schemes attempting to obtain nuclear
energy levels starting from a
cut-off independent low energy 2 body potential \cite{vlowk}.
In order to the test the validity
of this approximation it is useful to have a many fermion model which can
be exactly solved and in which the cluster approximation can be easily
implemented.
In what follows, we will construct both analytical and numerical
effective hamiltonians for the Lipkin Model \cite{Lipkin} and study the
effects of neglecting induced operators on the ground state energy.

It is important to note at the outset that due to
particular features of the Lipkin Model our implementation
of the cluster approximation does not
coincide with the truncation practised in \cite{assorted}.
In addition, we emphasize that the cluster approximation
makes the most sense when applied to systems of low-density and in our
Lipkin model application there is no length scale.  Equivalently,
the interactions have no limit to their range.  For both these reasons,
our conclusions concerning the limits of the cluster approximation
are restricted to the present application.  Nevertheless, the ability to
directly investigate the validity limits of the cluster approximation and the
dependence on the coupling strength impart significance to these results
in the long-standing tradition of testing many-body methods with
simple soluble examples.

We adopt the Lipkin Model for the case of $N $ spin 1/2 fermions in an 
external magnetic
field which interact with one another via couplings described by
parameters $v$ and $w$. The Hamiltonian may be written as follows
\begin{displaymath}
\frac{1}{2}\mu\sum (\sigma\;a^{\dagger}_{p\sigma}a_{p\sigma}) +
\frac{v}{2}\sum\;(a^{\dagger}_{p\sigma}a^{\dagger}_{p^{'}\sigma}
a_{p^{'}-\sigma}a_{p-\sigma}) +
\frac{w}{2}\sum\;(a^{\dagger}_{p\sigma}a^{\dagger}_{p^{'}-\sigma}
a_{p^{'}\sigma}a_{p-\sigma})
\end{displaymath} where the sum runs over the single particle states labelled 
by
$p$ and $p^{'}$ and $\sigma$ which takes values $\pm$ 1. The range of $p$ is 
taken as $1 \leq p \leq N$.
It is customary
to introduce quasi-spin operators \begin{eqnarray}
J_{+} & = & \sum_{p}a^{\dagger}_{p,+1}a_{p,-1} \\
J_{-} & = & \sum_{p}a^{\dagger}_{p,-1}a_{p,+1} \\
J_{z} & = & \frac{1}{2}\sum_{p\sigma}\sigma a^{\dagger}_{p\sigma}a_{p\sigma} \\
\end{eqnarray}
which satisfy the usual angular momentum commutation relations.  Hence,
we will use the term "angular momentum" to characterize these quasi-spin
operators.
In terms of the quasi-spin operators the
Hamiltonian for the Lipkin Model may be written (following \cite{Lipkin}) as
\begin{equation}
H = \mu\,J{_z} + \frac{1}{2}\,v\,(J_{+}^{2} + J_{-}^{2}) +
\frac{1}{2}\,w\,(J_{+}J_{-} + J_{-}J_{+}) 
\end{equation} 
\noindent after dropping a term proportional to $w$ which is of no consequence
as $w$ will be eventually set to zero.
A natural basis for defining the eigenstates of the hamiltonian is the
complete set of states of the total angular momentum operator and the
$z$ component of the total angular momentum {\em i.e.}
$ \vec{J} = \sum \vec{j_{p}}$ and $J_{z} = \sum j_{z_{p}}$
with $\vec{j_{p}}=\frac{1}{2}\vec{\sigma_p}$.
In our work, the description in terms of the total angular momentum is
equivalent to the description in terms of the quasi-spin.
Since $\left[H,J^{2}\right] = 0 $ we can
restrict our attention to multiplets with fixed $j$. We choose the multiplet
the largest value of $j$, thus
the number of particles is simply twice the
largest eigenvalue of $J_{z}$. 

For the sake of simplicity we will set $w = 0$ and also
add to the hamiltonian a constant term $v\,j(j+1) + \mu\,j$ for the purpose
of making
the hamiltonian positive definite, {\em i.e.} our hamiltonian in a fixed $j$
multiplet reads
\begin{equation}
H = \mu\,(J{_z} + J^{2}/(j+1)) + \frac{1}{2}\,v\,(J_{+}^{2} + J_{-}^{2}) +
v\,J^{2} \label{eq:hlip}
\end{equation}
In order to get a feeling for the nature of operators induced in the
effective hamiltonian we will consider the case where $\mu = 0$ and
$j = 4$. We label states according to their $J_{z}$ eigenvalues; since
there are no terms in the hamiltonian coupling states whose
$J_{z}$ eigenvalues are odd to those whose $J_{z}$ eigenvalues are even,
we will consider only states whose $J_{z}$ eigenvalues run over even values
between -4 to +4. The hamiltonian is thus five dimensional. Setting $v$ to 1,
the bare hamiltonian (in the basis -4,4,-2,0,2) takes the form
\begin{displaymath}
\left(
\begin{array}{ccccc}
20      & 0      & 5.2915 & 0      &   0   \\
0       & 20     & 0      & 0      & 5.2915\\
5.2915  & 0      & 20     & 9.4648 &   0   \\
0       & 0      & 9.4648 & 20     & 9.4648\\
0       & 5.2915 &  0     & 9.4648 & 20    \\
\end{array}
\right)
\end{displaymath}
Note that with $\mu = 0$ the hamiltonian is symmetric under
$J_{i} \rightarrow -J_{i}$.
Following \cite{leesuz} we define a $P$ Space to contain all states to be
treated exactly, and a Q space which is averaged over. We choose the P Space
to span all states with
$J_{z}$ eigenvalues running from -2 to +2 and the $Q$ space the remaining
two states.

In this basis PHP is
\begin{displaymath}
\left(
\begin{array}{ccc}
20     & 9.4648 &   0   \\
9.4648 & 20     & 9.4648\\
0      & 9.4648 & 20    \\
\end{array}
\right)
\end{displaymath}
QHQ is
\begin{displaymath}
\left(
\begin{array}{cc}
20     &   0 \\
0      & 20  \\
\end{array}
\right)
\end{displaymath}
and QHP is given by
\begin{displaymath}
\left(
\begin{array}{ccc}
5.2915 & 0      &   0   \\
0      & 0      & 5.2915\\
\end{array}
\right)
\end{displaymath}

The full hamiltonian may be schematically represented by
\begin{displaymath}
\left(
\begin{array}{cc}
\rm{QHQ} & \rm{QHP} \\
\rm{PHQ} & \rm{PHP} \\
\end{array}
\right)
\end{displaymath} which is the form which will be used throughout this paper.

In order to arrive at a non-hermitean form of the effective hamiltonian
we use the following iteration scheme due to Andreozzi \cite{ALS}:
\begin{eqnarray}
                 X_{0} & = & \frac{-1}{QHQ}QHP \\
                 X_{n} & = & \frac{1}{q_{n-1}}X_{n-1}p_{n-1} \\
                 p_{n-1} & = & PHP + PHQ\sigma_{n-1} \\
                 q_{n-1} & = & QHQ - \sigma_{n-1}PHQ
\end{eqnarray}
$p_{n}$ and $q_{n}$ are the P and Q space effective hamiltonians after $n$
iterations, where
$n = 1,2,3 \ldots$ \& $\sigma_{n-1} = X_{0} + X_{1} \ldots X_{n-1}$.
The iterations converge when successive
$X_{i}$ become smaller and smaller. Once convergence has been
attained $\omega = \sum X_{i}$
is a solution to the Lee Suzuki decoupling equation,
\begin{equation}
\omega PHQ \omega + \omega PHP - QHQ \omega - QHP = 0
\end{equation}
The iterations defined in Eqs.3-6 are stable and converge
smoothly due to the presence of the $v\,J^{2}$ term
added to the Lipkin Model Hamiltonian. The eigenvalues of the non-hermitean
effective hamiltonian are in excellent agreement with the lowest three
eigenvalues of the full hamiltonian.

In order to analyze the operators induced by the iteration
procedure it is necessary to transform the non-hermitean effective hamiltonian
to a hermitean form
\cite{Wilson}
\begin{equation}
H_{eff} = \frac{1}{\sqrt{I_{P} + \omega^{T}\omega}}
(I_{P} + \omega^{T})H(I_{P} + \omega)
\frac{1}{\sqrt{I_{P} + \omega^{T}\omega}}
\end{equation}
In more transparent notation, $H_{eff}$ can be expressed as
\begin{equation}
H_{eff} = \frac{1}{\sqrt{I_{P} + \omega^{T}\omega}}
(\omega^{T}QHQ\omega + \omega^{T}QHP + PHQ\omega + PHP )
\frac{1}{\sqrt{I_{P} + \omega^{T}\omega}}
\end{equation}
This form of the effective hamiltonian arises if the only transformations
made are those needed to decouple part of the Hilbert space with no
additional unitary transformations on the basis states \cite{Suzuki}.
This
prescription leads to the following hermitean effective hamiltonian
\begin{displaymath}
\left(
\begin{array}{ccc}
12.1103 & 4.4521  & -2.5982 \\
4.4521 & 16.0685 & 4.4521   \\
-2.5982& 4.4521  & 12.1103   \\
\end{array}
\right)
\end{displaymath}

We also evaluated the effective Hamiltonian using a
matrix inversion scheme \cite{leesuz,assorted} and observe that it
produces the same P-space hermitean matrix as the Andreozzi iteration
scheme that we use throughout this work.

Comparing the hermitean effective hamiltonian with PHP
it is evident that in addition to a global rescaling of
the bare matrix elements, operators consistent with the
symmetries of the bare hamiltonian and the cutoff but not present in the bare
hamiltonian have been induced.
The hermitean effective hamiltonian, $H_{eff}$, may also
be expressed analytically as
\begin{equation} H_{eff} =
{\rm P}(0.8033\,J^{2} -0.9889\,J_{z}^{2} +
0.2346(J_{+}^{2} + J_{-}^{2}) -\frac{0.1443}{j(j+1)}(J_{+}^{4} + J_{-}^{4})\; )
{\rm P}\label{eq:heff}
\end{equation}
where $\rm{P}$ is a projection operator which is included to ensure that
$H_{eff}$ has non-vanishing matrix elements only between P space
states.
$\rm{P}$ has the form \begin{displaymath}
\frac{J^{2}}{j(j+1)} + \gamma (J_{+}^{8}J_{-}^{8} + J_{-}^{8}J_{+}^{8})
\end{displaymath} The
constant $\gamma$ is given by
$-\left(<-4 \mid J_{-}^{8}\mid 4><-4 \mid J_{+}^{8} \mid 4> \right){}^{-1}$.
It is worth pointing out that in addition to new off diagonal operators,
diagonal operators not present in Eq. \ref{eq:hlip} appear in Eq.
\ref{eq:heff}.
Note that $<-2 \mid H_{eff}\mid -2 > = < 2 \mid H_{eff} \mid 2 >$,
which rules out the possibility of Eq. \ref{eq:heff} containing
an induced diagonal operator of the form $J_{z}^{m}$ ($m$ odd), which in any
case would not have been expected on symmetry grounds.

The new diagonal operators have the same effect as a state dependent
rescaling of the  diagonal matrix elements of the bare hamiltonian; a similar
feature arises in the nuclear many body problem.
In the cluster approximation, the matrix elements of the
effective hamiltonian are typically constrained to be related to those of the
bare hamiltonian by channel dependent renormalization. To implement the analog
of the cluster approximation
would require setting
$<2 \mid H_{eff}\mid -2>= <-2 \mid H_{eff}\mid +2>= 0$
or equivalently dropping the $ J_{\pm}^{4}$ terms in Eq. \ref{eq:heff}.
Since the dimension of the effective hamiltonian is rather small there are just
two matrix elements to be omitted in this two body cluster approximation.
With a larger value of $j$, or equivalently
a larger number of fermions, the number of matrix elements omitted in the
cluster approximation will be larger.
The rest of this paper is devoted to a detailed numerical investigation of
the validity of the cluster approximation in the Lipkin Model with larger
values of $j$ than we have hitherto considered.

We next consider the case where $j = 8$, $\mu = 0.1$ and $ v = 0.03$.
With these parameters the dimensionless quantity $(v\,j)$/$\mu$ is 2.4.
   Thus we are in a regime where perturbation theory cannot necessarily be
trusted.
The Q-Space is chosen to be one dimensional,
containing the state whose $J_{z}$ eigenvalue is +8. All remaining states are
absorbed into the P-space which is 8 dimensional as we once again keep only
states whose $J{_z}$ eigenvalues are even. Choosing the basis for H as
8,6,4,2,0,-2,-4,-6,-8 PHP is given by
\begin{displaymath}
\left(
\begin{array}{cccccccc}
3.5600  &0.7010  &0       &0      &0       &0       &0        &0      \\
0.7010  &3.3600  &0.9439  &0      &0       &0       &0        &0 \\
0       &0.9439  &3.1600  &1.0649 &0       &0       &0        &0  \\
0       &0       &1.0649  &2.9600 &1.0649  &0       &0        &0 \\
0       &0       &0       &1.0649 &2.7600  &0.9439  &0        &0  \\
0       &0       &0       &0      &0.9439  &2.5600  &0.7010   &0      \\
0       &0       &0       &0      &0       &0.7010  &2.3600   &0.3286 \\
0       &0       &0       &0      &0       &0       &0.3286   &2.1600 \\
\end{array}
\right)
\end{displaymath}

and QHP is given by
\begin{displaymath}
\left(
\begin{array}{cccccccc}
0.3286  &0  &0       &0       &0   &0   &0  &0  \\
\end{array}
\right)
\end{displaymath}
QHQ is simply $<8 \mid H \mid 8> $  or 3.7600
Once again the iterative scheme defined earlier is implemented leading to
the following hermitean effective hamiltonian, $H_{eff}$
\begin{displaymath}
\left(
\begin{array}{cccccccc}
   {\bf 3.3159} & {\bf 0.4251} & -0.3246
& -0.2780 & -0.1751 & -0.0782 & -0.0022 & -0.0027 \\
   {\bf 0.4251} & {\bf 3.1360} & {\bf 0.6805} & -0.2254 & -0.1418 & -0.0633
& -0.0180 & -0.0022 \\
-0.3246 & {\bf 0.6805} & {\bf 2.8501}  & {\bf 0.7997}  & -0.1668
& -0.0745 & -0.0211 & -0.0026 \\
-0.2780 & -0.2254 & {\bf 0.7997}  & {\bf 2.7331}  & {\bf 0.9221}  &  -0.0637
&-0.0181  & -0.0022 \\
-0.1751 & -0.1418& -0.1668 & {\bf 0.9221}&{\bf 2.6702}&{\bf 0.9038}&-0.0114
& -0.0014 \\
-0.0782 & -0.0633& -0.0745 & -0.0637 &
{\bf 0.9038}& {\bf 2.5421}&{\bf 0.6959}  &-0.0006 \\
-0.0222 &-0.0180 & -0.0211
& -0.0181 & -0.0114 &{\bf 0.6959} & {\bf 2.3586}&{\bf 0.3285} \\
-0.0027 & -0.0022& -0.0026 &  -0.0022 &-0.0014
&-0.0006  &{\bf 0.3285}&{\bf 2.1600} \\
\end{array}
\right)
\end{displaymath}
Comparing the $H_{eff}$ with PHP we see
immediately that new operators have been induced,
the induced transitions corresponding to
$\mid \Delta J_{z} \mid = 4,6,\ldots 14 $.
The lowest eigenvalue of the effective hamiltonian is 1.0460 which is in
excellent agreement with the lowest eigenvalue of the full hamiltonian.

It is noteworthy that the magnitude of matrix elements of
the induced operators is generally smaller than those of the operators
present in the bare hamiltonian despite the fact that the
interaction has for
all intents and purposes infinite range. This will be a recurring
feature of
the effective hamiltonians we construct and suggests that in hamiltonians with
a finite range interaction, such as the nuclear many body problem, the
relative size
of induced  many body operators may indeed be small.

Implementing the 2 body cluster
approximation corresponds to retaining all matrix
elements of $H_{eff}$ which are in bold face and setting all other matrix
elements to 0. The lowest eigenvalue of the hamiltonian which results is
1.1238, thus the effect of the cluster approximation is not too drastic.
In order to offset the possible effect of the shift in the ground state
energy, we consider the splitting between the ground and first excited
state as is done in the treatment of the nuclear many body problem
\cite{assorted}.
With no cluster truncation of $H_{eff}$ , we obtain 0.5974. With the
cluster truncation we get 0.5769; the effects of the cluster truncation are
apparently comparable in size and sign for both the ground and first excited
states.
This might have been anticipated on perturbative grounds alone, as the
state left out of the P space has the highest unperturbed energy and thus
might be expected to have a minimal effect on the true ground state energy.

As a check on this line of reasoning we
consider the case where the Q space consists just of the state
whose $J_{z}$ eigenvalue is 0. Since the unperturbed energy of the omitted
state is not as high as before, the cluster truncation estimate for
the ground state energy can be expected to be less accurate than before.
Furthermore, with this choice of Q space we can
address the efficacy of the cluster approximation when
intruder states are present. Strictly speaking intruder states arise when
the perturbed energies of Q space states lie among the perturbed energies of
the P space states. With our new choice of Q space, the unperturbed energies
of the Q space states lie among the unperturbed energies of the P space
states, simulating the effect of intruder states.
Apart from the change in the choice of Q space,
all other parameters retain the same values
as before.
Choosing the basis for H as 0,-8,-6,-4,-2,2,4,6,8 PHP is given by
\begin{displaymath}
\left(
\begin{array}{cccccccc}
2.1600  &0.3286  &0       &0      &0       &0       &0        &0      \\
0.3286  &2.3600  &0.7010  &0      &0       &0       &0        &0 \\
0       &0.7010  &2.5600  &0.9439 &0       &0       &0        &0  \\
0       &0       &0.9439  &2.7600 &0       &0       &0        &0 \\
0       &0       &0       &0      &3.1600  &0.9439  &0        &0  \\
0       &0       &0       &0      &0.9439  &3.3600  &0.7010   &0      \\
0       &0       &0       &0      &0       &0.7010  &3.5600   &0.3286 \\
0       &0       &0       &0      &0       &0       &0.3286   &3.7600 \\
\end{array}
\right)
\end{displaymath}
In the same basis QHP is given by
\begin{displaymath}
\left(
\begin{array}{cccccccc}
0&  0&  0& 1.0649& 1.0649   &0   &0  &0  \\
\end{array}
\right)
\end{displaymath}
   and QHQ is just $< 0 \mid H \mid  0 > $ or 2.9600.
Repeating the same iterative procedure yields the following $H_{eff}$,
\begin{displaymath}
\left(
\begin{array}{cccccccc}
{\bf 2.1600} & {\bf 0.3285} &
-0.0005 & -0.0052 & -0.0063 & -0.0020 & -0.0011 & -0.0003 \\
{\bf 0.3285} & {\bf 2.3588} & {\bf 0.6966} &
-0.0403 & -0.0490 & -0.0158 & -0.0092  & -0.0027 \\
-0.0005 &
{\bf 0.6966} & {\bf 2.5443} & {\bf 0.8072}&
-0.1675 & -0.0564 & -0.0328 & -0.0097 \\
-0.0052 & -0.0403 &
{\bf 0.8072} & {\bf 2.2399} &
-0.7796 & -0.4809 & -0.2771 & -0.0792 \\
-0.0063 & -0.0490 & -0.1675 & -0.7796 &
{\bf 2.0606} & {\bf 0.3524} &
-0.3415 & -0.0981 \\
-0.0020 & -0.0158 & -0.0564 & -0.4809 &
{\bf 0.3524} & {\bf 3.1571} & {\bf 0.5830} &
-0.0347 \\
-0.0011 & -0.0092 & -0.0328 & -0.2771 & -0.3415 &
{\bf 0.5830} & {\bf 3.4914} & {\bf 0.3085} \\
-0.0003 & -0.0027 & -0.0097 & -0.0792 & -0.0981 & -0.0347 &
{\bf 0.3085} & {\bf 3.7541} \\
\end{array}
\right)
\end{displaymath}
Once again, the lowest eigenvalue of $H_{eff }$ coincides with that of the
full Hamiltonian to 5 significant figures. In order to implement the two body
cluster approximation, we retain only the bold faced matrix elements of
$H_{eff}$, setting all other matrix elements to 0. The lowest eigenvalue of
the resulting hamiltonian overshoots the true ground state by 25\%.
On the other hand, the splitting between
the first excited state and the
ground state with the cluster truncation is 0.6228, compared with the 
true value of 0.5974.
Thus, the accuracy of the cluster approximation depends on the choice of
the observable: the absolute eigenvalue or the eigenvalue spacing. 

As a further check we now consider a situation where
{\em i.e.} $\rm{QHQ} = <-8 \mid H \mid -8 > $ keeping all the coupling
constants the same. The unperturbed Q space energy is smallest
among all the cases we have considered; the effect of the cluster truncation
on the ground state energy may be expected to be the largest.
In the basis in which the Hamiltonian is ordered
-8,-6,-4,-2,0,2,4,6,8 PHP takes the form
\begin{displaymath}
\left(
\begin{array}{cccccccc}
2.3600  &0.7010  &0       &0      &0       &0       &0        &0      \\
0.7010  &2.5600  &0.9439  &0      &0       &0       &0        &0 \\
0       &0.9439  &2.7600  &1.0469 &0       &0       &0        &0  \\
0       &0       &1.0469  &2.9600 &1.0649  &0       &0        &0 \\
0       &0       &0       &1.0649 &3.1600  &0.9439  &0        &0  \\
0       &0       &0       &0      &0.9439  &3.3600  &0.7010   &0      \\
0       &0       &0       &0      &0       &0.7010  &3.5600   &0.3286 \\
0       &0       &0       &0      &0       &0       &0.3286   &3.7600 \\
\end{array}
\right)
\end{displaymath}
and QHQ is given by
\begin{displaymath}
\left(
\begin{array}{cccccccc}
0.3286 &0 &0 &0 &0 &0 &0 &0
\end{array}
\right)
\end{displaymath}

In exactly the same manner as before, we compute the effective hamiltonian and
obtain
\begin{displaymath}
\left(
\begin{array}{cccccccc}
{\bf 2.3315} & {\bf 0.6434} &
-0.1302 & -0.2085 & -0.2453 & -0.2098 & -0.1215 & -0.0360 \\
{\bf 0.6434} & {\bf 2.5111} & {\bf 0.8336} &
-0.1764 & -0.2072 & -0.1771 & -0.1025  & -0.0304 \\
-0.1302 &
{\bf 0.8336} & {\bf 2.5112} & {\bf 0.6671}&
-0.4672 & -0.3993 & -0.2311 & -0.0686 \\
-0.2085 & -0.1764 &
{\bf 0.6671} & {\bf 2.3240} &
{\bf 0.3179} & -0.6383 & -0.3695 & -0.1096 \\
-0.2453 & -0.2072 & -0.4672 & {\bf 0.3179} &
{\bf 2.2826} & {\bf 0.1942} &
-0.4340 & -0.1288 \\
-0.2098 & -0.1771 & -0.3993 & -0.6383 &
{\bf 0.1942} & {\bf 2.7193} & {\bf 0.3301} &
-0.1100 \\
-0.1215 & -0.1025 & -0.2311 & -0.3695 & -0.4340 &
{\bf 0.3301} & {\bf 3.3453} & {\bf 0.2649} \\
-0.0360 & -0.0304 & -0.0686 & -0.1096 & -0.1288 & -0.1100 &
{\bf 0.2649} & {\bf 3.7411} \\
\end{array}
\right)
\end{displaymath}
Retaining just the bold faced terms in the two body cluster
approximation leads to
a 20 \% error in the estimate of the ground state energy.
Furthermore, the splitting between the first excited state and the ground state
is now 0.6293, compared with the true value of 0.5974.
As expected the
error is substantial but has actually decreased instead of increasing as
we originally anticipated on the basis of perturbative arguments.
Our results are summarised in the table below:
($E_{gs}$ is the ground state energy and 0.5974 is the true splitting between
the ground state and the first excited state)
\newline
\vspace{.2cm}
\newline
\begin{tabular}{|l|l|l|l|l|}\hline
Q Space $J_{z}$ & $< Q \mid H \mid Q> $ & $E_{gs}$
($H_{eff}$) & $E_{gs}$ (Cluster) & Splitting \\ \hline
8  & 3.7600 & 1.0460 & 1.1238 & 0.5769 (0.5974) \\ \hline
0  & 2.9600 & 1.0460 & 1.3130 & 0.6228 (0.5974) \\ \hline
-8 & 2.1600 & 1.0460 & 1.2547 & 0.6293 (0.5974)\\ \hline
\end{tabular}
\newline
\vspace{.2cm}
\newline
Comparing
the three cases we may conclude that even when the dimensions of the P and Q
spaces are fixed,
the accuracy of the cluster
approximation in fixing the ground state 
is strongly affected by the choice of P and Q spaces in a
manner which cannot be anticipated by perturbation theory alone. The 
splitting between the first excited state and the ground state is rather less 
sensitive to the choice of P space.
As might be
expected, larger cluster sizes can improve the estimate of the ground state
energy. In the case where the Q Space contains $<-8 \mid H \mid -8>$, going
from 2 body to 4 body
clusters reduces the discrepancy in the ground state energy from 20\% to
15\% which is an appreciable but not spectacular improvement.

We now investigate what happens when the size of the P space
relative to the full Hilbert space shrinks;
the Q Space is spanned by states whose eigenvalues under
$J_{z}$ are 8,6,-6, \& 8 leaving a 5 dimensional P space.
If the basis for the Hamiltonian is ordered 8,6,-6,-8,-4,-2,0,2,4 QHQ is
given by
\begin{displaymath}
\left(
\begin{array}{cccc}
3.7600 & 0.3268 & 0      & 0 \\
0.3286 & 3.5600 & 0      & 0 \\
0      & 0      & 2.3600 & 0.3286 \\
0      & 0      & 0.3286 & 2.1600 \\
\end{array}
\right)
\end{displaymath}
PHP is given by
\begin{displaymath}
\left(
\begin{array}{ccccc}
2.5600 & 0.9439 & 0      & 0      & 0     \\
0.9439 & 2.7600 & 1.0649 & 0      & 0     \\
0      & 1.0649 & 2.9600 & 1.0649 & 0     \\
0      & 0      & 1.0649 & 3.1600 & 0.9439 \\
0      & 0      & 0      & 0.9439 & 3.3600 \\
\end{array}
\right)
\end{displaymath}
and QHP by
\begin{displaymath}
\left(
\begin{array}{ccccc}
0      & 0      & 0      & 0      & 0     \\
0      & 0      & 0      & 0      & 0.7010  \\
0.7010 & 0      &0       & 0      & 0     \\
0      & 0      & 0      & 0      & 0     \\
\end{array}
\right)
\end{displaymath}

Once again we construct the effective Hamiltonian by iteration and $H_{eff}$ is
obtained to be
\begin{displaymath}
\left(
\begin{array}{ccccc}
{\bf 1.5688}  & {\bf 0.1692} & -0.4703 & -0.0242 & 0.0993     \\
{\bf 0.1692}  & {\bf 2.0703} & {\bf 0.3397}  & -0.4884 & -0.1198  \\
-0.4703 & {\bf 0.3397} &{\bf 1.9539}& {\bf 0.1502} & -0.3723     \\
-0.0242 & -0.4884 &{\bf 0.1502}& {\bf 2.1449}  & {\bf 0.3063}   \\
0.0993  & -0.1198&-0.3723  &{\bf 0.3063}   &{\bf 2.4330}   \\
\end{array}
\right)
\end{displaymath}
Once again, the lowest eigenvalue of $H_{eff}$ and that of the full hamiltonian
agree to 5 significant figures. Implementing the cluster approximation amounts
to retaining only the bold faced terms in $H_{eff}$. Doing so yields a
truncated effective Hamiltonian whose lowest eigenvalue overestimates the
true ground state energy by over 41\%. In addition, the splitting between the
ground state energy and the first excited state is 0.2134, drastically
different from the true value of 0.5974.
This suggests that the accuracy of the
two body cluster approximation decreases as the size of the P space
relative to the full Hilbert space decreases.

Finally, we investigate what happens when the coupling strength $v$ which was
hitherto fixed to be 0.03 is allowed to vary.
We go back to the case where the cluster approximation gave the best results,
{\em i.e.} with
$ \rm{QHQ} = < 8 \mid H \mid 8> $, but now
set $v$ to be 0.1.
Once again, $H_{eff}$ with no cluster truncation
reproduces the true ground state energy to five significant figures. However,
if the two body cluster approximation is implemented on $H_{eff}$, there is a
39 \% error in the ground state energy instead of just under 8 \% with the
original value of $v$. In addition the splitting between the ground state and
the first excited state is poorly reproduced in the cluster approximation.

As expected, the accuracy of the two body cluster approximation decreases at
large couplings.

To conclude, we have investigated the relevance of induced operators which
arise in an effective operator treatment of the Lipkin Model. Ignoring
these operators as is done in the cluster approximation, leads to
errors of varying magnitude
but which can be uncomfortably large. The errors
from the cluster approximation in these soluble examples are likely to be
larger than those in more realistic applications due
in part due to the infinite range of the interaction in the Lipkin Model.
In hamiltonians where the interaction is of finite range, such as the
nuclear many body problem the errors may not be as large. Nonetheless,
even with infinite range interactions, it is possible to establish guidelines
for minimising the error which may well be relevant for a broader class of
interactions. Based on our numerical results,
the approximation seems to work best for couplings which are
not too large and when the model space states constitute as large a
fraction as possible of the full Hilbert space.
In other instances the errors can be
substantial even when the couplings are only moderately large.  We note
that more realistic applications typically investigate carefully
the convergence rate of the eigenvalues, at fixed cluster size, with
increasing P-space dimensions.  In this way, one anticipates that the
errors due to cluster truncation are minimized.

This work was supported in part by U.S. DOE Grant No. DE-FG-02-87ER-40371,
Division of High Energy and Nuclear Physics.

\end{document}